\documentstyle[preprint,aps]{revtex}
\begin{document}
\draft

\title{\bf
Critical Finite-Size-Scaling Amplitudes of a Fully Anisotropic
Three-Dimensional Ising Model}
\author{M.~A.~Yurishchev}
\address{
Vasilsursk Laboratory, Radiophysical Research Institute,
606263 Vasilsursk, Nizhny Novgorod Region, Russia}

\date{\today}
\maketitle

\begin{abstract}
A fully anisotropic simple-cubic Ising lattice in the geometry
of periodic cylinders $n\times n\times\infty$ is investigated by
the transfer-matrix finite-size scaling method.
In addition to the previously obtained critical amplitudes of the
inverse correlation lengths and singular part of the free energy
[M.~A.~Yurishchev, Phys. Rev. B {\bf 50}, $13\,533$ (1994)],
the amplitudes of the usual (``linear'') and
nonlinear susceptibilities and the amplitude of the second
derivative of the spin-spin inverse correlation length with
respect to the external field are calculated.
The behavior of critical amplitude combinations (which, in accordance
with the Privman-Fisher equations, do not contain in their
composition the nonuniversal metric coefficients and geometry
prefactor) are studied as a function of the interaction anisotropy
parameters.
A universality domain for the amplitude ratios is found in the
quasi-one-dimensional regime of interactions in the system.
In the case of a fully {\it isotropic\/} three-dimensional Ising
model for which the high precision values of the critical coupling
and critical-point free energy are available, improved estimates
are obtained for the following four universal quantities:
(1) the amplitude of spin-spin inverse correlation length,
(2) the amplitude of singular part of the free energy,
(3) the ratio of the amplitude of a second derivative of the
spin-spin inverse correlation length with respect to the external
field to the usual susceptibility amplitude, and
(4) the ratio of the nonlinear susceptibility amplitude to the
square of the linear susceptibility amplitude (i.e., for the
finite-size counterpart of the four-point renormalized coupling
constant).
\end{abstract}



\section{Introduction}
\label{sec:Intro}
In paper \cite{Yu94} (hereafter referred to as I), the critical
finite-size amplitudes of the inverse correlation lengths and
the ``singular'' part of the free energy per spin have been
calculated for the three-dimensional ferromagnetic Ising model on
a simple-cubic lattice with fully anisotropic interactions.
The amplitudes have been obtained by a finite-size scaling (FSS)
analysis of the transfer matrix (TM) data for subsystems having
a shape of $n\times n\times\infty$ cylinders with periodic boundary
conditions in both transverse directions.

The results showed that the ratios of the above critical amplitudes
practically do not depend on the anisotropy parameter $J_x/J_z$ if
the second parameter $J_y/J_x$ is fixed (the parallelepipeds
$n\times n\times\infty$ are assumed to be stretched along
$z$ axis; $J_x$ and $J_y$ are the interaction constants in the
transverse directions of a cylinder, and $J_z$ is the interaction
constant in the longitudinal direction).
Furthermore, it has been found that the amplitude ratios are
independent of the second anisotropy parameter, $J_y/J_x$, in the
vicinity of $J_y/J_x=1$.
Unfortunately, a wrong conclusion concerning the amplitude ratio
constancy in the full range of the $J_y/J_x$ values has been made
in I.

In addition to the amplitude of the spin-spin inverse correlation
length ($A_s$) and the free-energy amplitude ($A_f$), in the present
paper we calculate the amplitudes of linear and nonlinear
susceptibilities ($A_{\chi}$ and $A_{\chi^{(4)}}$, respectively), as
well as the amplitude of the second derivative of the spin-spin
inverse correlation length with respect to the external field
($A_{\kappa^{\prime\prime}_{hh}}$).
The calculations are performed again for the cyclic clusters
$n\times n\times\infty$ with the maximum number of lattice layers
in each transverse direction $n=4$ (i.e., for the sizes of TMs up to
$65\,536\times65\,536$).
Guiding ourselves by the functional equations of Privman and Fisher
\cite{PF,PPHA,ChH}, we construct from critical amplitudes the
combinations $A_{\kappa^{\prime\prime}_{hh}}/A_{\chi}$ and
$A_{\chi^{(4)}}A_s/A^2_{\chi}$ which, as $A_s/A_f$, do not contain
the certain nonuniversal coefficients.
The analysis shows that those amplitude ratios are universal with
respect to the anisotropy parameter $J_x/J_z\to 0$ but depend on
the parameter $J_y/J_x$.
However, the $J_y/J_x$ dependencies exhibit smooth extrema near
$J_y/J_x=1$, and therefore the amplitude ratios have a local
universality in their vicinity upon the second anisotropy
parameter of the model.

In the present paper we find also the values of $A_s$, $A_f$,
$A_{\kappa^{\prime\prime}_{hh}}/A_{\chi}$, and
$A_{\chi^{(4)}}/A^2_{\chi}$ for $n=2, 3$, and 4 in the special case
of the fully isotropic simple-cubic Ising lattice by using the
available high accurate estimates for the critical point and critical
free energy and give three-point extrapolations for these quantities.


\section{Spatial Anisotropy and the Privman-Fisher Equations}
\label{sec:SAPFE}
In the case of $d$-dimensional cylinders with sizes of
$L_1\times L_2\times\cdots\times L_{d-1}\times\infty$, the
Privman-Fisher equations which allow to identify for the
hyperscaling systems the universal combinations of critical
finite-size amplitudes have the form \cite{PF}
\begin{equation}
   \label{eq:kappa_tilde}
   \tilde{\kappa}(t, h)
   = L_0^{-1}X(C_1tL_0^{y_T}, C_2hL_0^{y_h}; L_1/L_0, \ldots ,
     L_{d-1}/L_0)
\end{equation}
and
\begin{equation}
   \label{eq:f_tilde}
   \tilde{f}^{(s)}(t, h)
   = L_0^{-d}Y(C_1tL_0^{y_T}, C_2hL_0^{y_h}; L_1/L_0,\ldots,
   L_{d-1}/L_0)\;.
\end{equation}
In these equations, $\tilde{\kappa}$ is the inverse correlation
length in the longitudinal direction of the cylinder,
$\tilde{f}^{(s)}$ is the singular part of the free-energy density
measured in units of $-k_BT$, $t=(T-T_c)/T_c$ is the reduced
temperature, $h$ is the normalized external field, $y_T$ and $y_h$ are
critical exponents, $C_1$ and $C_2$ are system-dependent metric
factors, and $L_0$ is a scale length.
The FSS functions $X(x_1, x_2; l_1, l_2,\ldots)$ and
$Y(x_1, x_2; l_1, l_2,\ldots)$ are universal (within the limits of
a given universality class) but may depend on the type of boundary
conditions and, as we see, on the aspect ratios $l_i=L_i/L_0$.
For the scaling length $L_0$, one can take
$L_0=(L_1L_2\ldots L_{d-1})^{1/(d-1)}$ or put
$L_0=\min(L_1,L_2,\ldots,L_{d-1})$, etc.

Let us pass in Eqs.\,(\ref{eq:kappa_tilde}) and (\ref{eq:f_tilde}) to
the dimensionless quantities which are immediately defined by the
parameters of a subsystem Hamiltonian.
Let the elementary cell of a simple-cubic lattice have the sizes of
$a_1\times a_2\times\cdots\times a_d$ and let the subsystem ``frame''
be a cylinder $n_1\times n_2\times\cdots\times n_{d-1}\times\infty$,
where $n_i$ is a number of spins in the $i$th transverse direction.
We will, however, restrict ourselves to the case of cylinders with
{\it square\/} cross section, i.e.,
$n_1=n_2=\ldots =n_{d-1}=n$.
Then $L_i=na_i$, $\tilde{\kappa}=a_d^{-1}\kappa_n$, and
$\tilde{f}^{(s)}=(a_1a_2\cdots a_d)^{-1}f_n^{(s)}$, where $\kappa_n$
and $f_n^{(s)}$ are, respectively, the dimensionless inverse
correlation length and singular part of the free energy per site.
(Below, however, we will often omit, for brevity, the word
``dimensionless''.)
In the new variables the Privman-Fisher equations can be written as
\begin{equation}
   \label{eq:kappa_n}
   \kappa_n(t, h)
   = n^{-1}G X(C_1tn^{y_T}, C_2hn^{y_h}; a_2/a_1,\ldots ,
     a_{d-1}/a_1)
\end{equation}
and
\begin{equation}
   \label{eq:f_n}
   f_n^{(s)}(t, h)
   = n^{-d}G\,Y(C_1tn^{y_T}, C_2hn^{y_h}; a_2/a_1,\ldots,
     a_{d-1}/a_1)\;,
\end{equation}
where $G=G(a_2/a_1,\ldots,a_{d-1}/a_1, a_d/a_1)$ is the
geometry prefactor.
We suppose that at least two lattice spacings in the system
are finite and nonzero; without loss of generality, we take them as
$a_1$ and $a_d$.

One obtains from Eqs.\,(\ref{eq:kappa_n})
and (\ref{eq:f_n}) that
\begin{equation}
   \label{eq:A_s}
   A_s=n\kappa_n(0, 0)=G X(0, 0; a_2/a_1, \ldots, a_{d-1}/a_1)\;,
\end{equation}
\begin{equation}
   \label{eq:A_f}
   A_f=n^df_n^{(s)}(0, 0)=G\,Y(0, 0; a_2/a_1, \ldots, a_{d-1}/a_1)\;,
\end{equation}
and analogously for the dimensionless critical finite-size-scaling
amplitudes of the derivatives of $\kappa_n(t,h)$ or $f_n^{(s)}(t,h)$.

The geometry-prefactor form depends on the choice of $L_0$.
The prefactor is defined with exactness up to some multiplicative
function of $a_2/a_1,\ldots,a_{d-1}/a_1$, inasmuch as such a function
can be introduced into the scaling functions $X$ and $Y$ or, vice
versa, taken out from them.
We will however put $G=1$ for the fully isotropic model.

In order to be able to employ the FSS theory, the ``lattice
spacings'' $a_i$ must be chosen so that in the vicinity of the phase
transition point the bulk correlation lengths
$\tilde{\xi}_1,\ldots,\tilde{\xi}_d$ along all the different spatial
directions become equal between themselves:
$\tilde{\xi}_1=\tilde{\xi}_2=\ldots=\tilde{\xi}_d$.
Such a rescaling of lattice spacings can be made for
systems with the isotropic critical exponent  $\nu$ of the bulk
correlation lengths, i.e.\ when $\xi_i=\xi_0^{(i)}t^{-\nu}$
(Refs.\cite{NB,BPP,BW}).
The choice of $a_i$ completes, in principle, the process of
expressing the quantities entering into the Privman-Fisher
equations via ``microscopic'' parameters --- the interaction
constants of the Hamiltonian.
Note that the situation is paradoxical to a certain extent: in order
to use the FSS theory for extracting information about a bulk
system from properties of its finite subsystems, one must know
before the correlation-length amplitudes of the bulk system.

In the two-dimensional anisotropic Ising lattice which is a limited
case ($J_y=0$) of the three-dimensional Ising model under
question, the dimensionless bulk correlation lengths for
$T>T_c$ are \cite{O} (see also, e.g., Ref.\cite{BG})
\begin{equation}
   \label{eq:xi_2D}
   \xi_x(J_x/k_BT,J_z/k_BT) = \xi_z(J_z/k_BT,J_x/k_BT)
   = 1/\ln[\coth(J_x/k_BT)\exp(-2J_z/k_BT)]\;.
\end{equation}
(When $T<T_c$, the correlation length expressions differ from
Eq.\,(\ref{eq:xi_2D}) by the factor $-1/2$ which is, however,
unessential for the present considerations.)
According to the isotropy requirement, $a_x\xi_x=a_z\xi_z$ and
therefore for the square Ising lattice
\cite{NB} (see also \cite{BPP,INW,KB})
\begin{equation}
   \label{eq:gamma_2D}
   G = \frac{a_z}{a_x} = \lim_{T\to T_c}\frac{\xi_x}{\xi_z}
   = \left[\sinh\left(\frac{2J_z}{k_BT_c}\right)\right]^{-1}\;,
\end{equation}
where the critical temperature $T_c$ is determined by the equation
\begin{equation}
   \label{eq:Tc2D}
   \sinh\left(\frac{2J_x}{k_BT_c}\right)
   \sinh\left(\frac{2J_z}{k_BT_c}\right)=1\;.
\end{equation}
Eqs.\,(\ref{eq:gamma_2D}) and (\ref{eq:Tc2D}) in parametric form
yield the geometry factor $G(J_x/J_z)$.
Note that up to date the anisotropy factors have been found for many
exactly solved two-dimensional models of statistical physics
\cite{KP,F,NK}.

In the three-dimensional space, Eqs.\,(\ref{eq:kappa_n}) and
(\ref{eq:f_n}) are written as:
\begin{equation}
   \label{eq:kappa_3D}
   \kappa_n(t, h)
   = n^{-1}G X(C_1tn^{y_T}, C_2hn^{y_h}; a_y/a_x)
\end{equation}
and
\begin{equation}
   \label{eq:f_3D}
   f_n^{(s)}(t, h)
   = n^{-3}G\,Y(C_1tn^{y_T}, C_2hn^{y_h}; a_y/a_x)\;.
\end{equation}
Due to the shape parameter $a_y/a_x$, the scaling functions
depend on the coupling ratios, and therefore
Eqs.\,(\ref{eq:kappa_3D}) and (\ref{eq:f_3D}) do not allow the
existence of any universal amplitude combinations.
An obvious exception, however, is given by the case when the
interactions in the transverse directions of the parallelepiped
$n\times n\times\infty$ are equal between themselves.
On physical grounds, it is clear that $a_x=a_y$ by this, and
consequently the critical amplitude combinations which do not
contain $C_1$, $C_2$, and $G$ will not depend upon the interaction
constants in the system.


\section{Formulas for the Susceptibilities and for the
Inverse Correlation-Length Derivative}
\label{sec:FSD}
For the amplitude ratio $A_{\kappa^{''}_{hh}}/A_\chi$, it follows
from Eqs.\,(\ref{eq:kappa_n}) and (\ref{eq:f_n}) that
\begin{equation}
   \label{eq:A1}
   \frac{A_{\kappa^{''}_{hh}}}{A_\chi}
   = \frac{\kappa^{''}_{hh}}{n^{d-1}\chi_n}\;,
\end{equation}
where $\kappa^{''}_{hh}=\partial^2\kappa_n/\partial h^2$, while
$\chi_n=\partial^2f_n/\partial h^2=\partial^2f^{(s)}_n/\partial h^2$
is the susceptibility of the system ($f_n$ denotes the dimensionless
free energy per lattice site).
Taking into account that $A_s=n\kappa_n$, one finds for the second
amplitude combination
\begin{equation}
   \label{eq:A2}
   \frac{A_{\chi^{(4)}}A_s}{A^2_\chi}
   = \frac{\chi^{(4)}_n\kappa_n}{n^{d-1}\chi^2_n}\;,
\end{equation}
where $\chi^{(4)}_n=-\partial^4f_n/\partial h^4$ is a nonlinear
susceptibility.
Thus, the problem is to carry out the calculation of
$\kappa^{''}_{hh}$ and $\chi^{(4)}_n$ for periodic cylinders
$n\times n\times\infty$ at $h=0$ since the method of calculating
$\kappa_n$ and $\chi_n$ was already done in I.

To solve these problems, we use, as in I, the TM technique.
The matrix elements of TM, $\cal{V}$, are given by
\begin{eqnarray}
   \label{eq:V}
   \langle S_{11},S_{12},\ldots,S_{nn}|{\cal V}|& &
   S_{11}^\prime,S_{12}^\prime,\ldots,S_{nn}^\prime
   \rangle=\prod_{i,j=1}^n
   \exp[\case{1}{2}K_x(S_{ij}S_{i+1j}
   +S_{ij}^\prime S_{i+1j}^\prime)\nonumber\\
   & &
   +\case{1}{2}K_y(S_{ij}S_{ij+1}+
   S_{ij}^\prime S_{ij+1}^\prime)
   +K_zS_{ij}S_{ij}^\prime+\case{1}{2}h(S_{ij}+S^{'}_{ij})]\;.
\end{eqnarray}
Here the spin variables $S_{ij}$ take the values $\pm1$;
$S_{in+1}=S_{i1}$ and $S_{n+1j}=S_{1j}$ for all $i,j=1,2,\ldots,n$;
$K_\alpha = J_\alpha/k_BT$ ($\alpha = x, y, z$).
The matrix $\cal{V}$ is real, symmetric, and dense; having all its
elements positive.
The dimensionless free energy per spin equals
\begin{equation}
   \label{eq:fn}
   f_n=\frac{1}{n^2}\ln\Lambda_1\;,
\end{equation}
where $\Lambda_1$ is the largest eigenvalue of $\cal{V}$.
Further, the dimensionless spin-spin inverse correlation length in
the longitudinal direction of a parallelepiped
$n\times n\times\infty$ is
\begin{equation}
   \label{eq:kapn}
   \kappa_n=\ln(\Lambda_1/\Lambda_2)\;,
\end{equation}
where $\Lambda_2$ denotes the second largest eigenvalue of the matrix
$\cal{V}$.

In order to derive the exact formulas for the zero-field derivatives
of the free energy and inverse correlation length, we will use
perturbation theory.
For this one expands the TM in powers of $h$:
\begin{equation}
   \label{eq:VV}
   {\cal V}=V+hV_1+h^2V_2+h^3V_3+h^4V_4 +O(h^5)\;.
\end{equation}

Let us use the symmetry of the model (in this context see I).
$\cal{V}$ is invariant under the transformations of the
group ${\bf T}\wedge{\bf C}_{2v}$ (${\bf T}$ is a group of
translations in the transverse directions of a cyclic bar
$n\times n\times\infty$ and ${\bf C}_{2v}$ is the point group
generated by two symmetry planes going through the middles of
opposite faces of the system).
Pass in Eq.\,(\ref{eq:VV}) to the basis of the identity irreducible
representation of the group ${\bf T}\wedge{\bf C}_{2v}$.
Expansion (\ref{eq:VV}) preserves the above form but now its terms
are blocks corresponding to the indicated representation.
Both eigenvalues $\Lambda_1$ and $\Lambda_2$ lie in the given block
$\cal{V}$.
Take now into consideration the symmetry ${\bf Z}_2$ (a group
of spin inversions).
The matrices $V$, $V_2$, and $V_4$ are symmetrical and the
matrices $V_1$ and $V_3$ are antisymmetrical (i.e., they change
sign) under the spin inversion operation.
Going by means of a similarity transformation into a new basis in
which the original representation of the group is completely
reducible, one obtains from Eq.\,(\ref{eq:VV})
\begin{eqnarray}
   \label{eq:tilde_V}
   \tilde{\cal V}
   =& &\left(
      \begin{array}{c|c}
         V^{(1)} & 0 \\ \hline
         0       & V^{(2)}
      \end{array}
   \right)
   +h\left(
      \begin{array}{c|c}
         0          & V^{(12)}_1 \\ \hline
         V^{(21)}_1 & 0
      \end{array}
   \right)
   +h^2\left(
      \begin{array}{c|c}
         V_2^{(1)} & 0 \\ \hline
         0         & V_2^{(2)}
      \end{array}
   \right)\nonumber\\ \nonumber\\
   & &
   +h^3\left(
      \begin{array}{c|c}
         0          & V^{(12)}_3 \\ \hline
         V^{(21)}_3 & 0
      \end{array}
   \right)
   +h^4\left(
      \begin{array}{c|c}
         V_4^{(1)} & 0 \\ \hline
         0         & V_4^{(2)}
      \end{array}
   \right)
   +O(h^5)\;.
\end{eqnarray}
For definiteness we suppose that the subblocks $V^{(1)}$,
$V_2^{(1)}$, and $V_4^{(1)}$ correspond to the identity irreducible
representation of the group ${\bf Z}_2$.
Denote the sizes of these subblocks by $N_1\times N_1$ and the
sizes of the subblocks $V^{(2)}$, $V_2^{(2)}$, and $V_4^{(2)}$ by
$N_2\times N_2$.
Then, the matrices $V_1^{(12)}$ and $V_3^{(12)}$ have the sizes
of $N_1\times N_2$; $V_1^{(21)}$ and $V_3^{(21)}$ are found by
transposing $V_1^{(12)}$ and $V_3^{(12)}$, respectively.
The numbers $N_1$ and $N_2$ can be obtained from a group-theoretical
analysis.
For the $2\times2\times\infty$ cluster, $N_1=5$ and $N_2=2$
(Ref.\cite{Yu93}); for the $3\times3\times\infty$ one, $N_1=N_2=18$
(Ref.\cite{Yu93}); and for $4\times4\times\infty$, $N_1=787$ and
$N_2=672$ (Ref.\ I).
Let, further, $\lambda_i^{(1)}$ and $\psi_i$ be the eigenvalues and
corresponding eigenvectors of matrix $V^{(1)}$; let
$\lambda_1^{(1)}=\lambda_1$ be the largest eigenvalue of the block
$V^{(1)}$ and consequently of the ``nonperturbed'' transfer matrix
$V$.
Due to the Perron theorem, $\lambda_1$ is nondegenerate.
Finally, let $\lambda_i^{(2)}$ and $\varphi_i$ be the eigenpair of
$V^{(2)}$; $\lambda_1^{(2)}$ is the largest eigenvalue of block
$V^{(2)}$.
Note that $\lambda_1^{(2)}$ is also nondegenerate if we do
not take the extreme cases which can be examined separately.

Using the stationary perturbation theory for a nondegenerate
level, we find the largest eigenvalue of $\cal{V}$ with accuracy up
to the terms of second order in $h$:
\begin{equation}
   \label{eq:L1}
   \Lambda_1=\lambda_1
   +\left[\psi_1^{+}V_2^{(1)}\psi_1
   +\sum_{k=1}^{N_2}\frac{(\psi_1^{+}V_1^{(12)}\varphi_k)^2}{\lambda_1
   -\lambda_k^{(2)}}\right]h^2+O(h^4)\;.
\end{equation}
From here, the expression for the initial (zero-field) susceptibility
follows as
\begin{equation}
   \label{eq:chi2}
   \chi_n=\frac{2}{n^2\lambda_1}\left[\psi_1^{+}V_2^{(1)}\psi_1
   +\sum_{k=1}^{N_2}\frac{(\psi_1^{+}V_1^{(12)}\varphi_k)^2}{\lambda_1
   -\lambda_k^{(2)}}\right]\;.
\end{equation}
The given formula, obtained by the perturbation-theory method,
differs in a form from the one derived in I using the
fluctuation-dissipation relation, but it is equivalent to it and
yields the same values of the susceptibility.

In an analogous way, one obtains the following expression for the
second largest eigenvalue of ${\cal V}$:
\begin{equation}
   \label{eq:L2}
   \Lambda_2=\lambda_1^{(2)}
   +\left[\varphi_1^{+}V_2^{(2)}\varphi_1
   +\sum_{k=1}^{N_1}\frac{(\psi_k^{+}V_1^{(12)}\varphi_1)^2}
   {\lambda_1^{(2)}-\lambda_k^{(1)}}\right]h^2+O(h^4)\;.
\end{equation}
From Eqs.\,(\ref{eq:kapn}), (\ref{eq:L1}), and (\ref{eq:L2}) we get
the work formula for the second derivative of the spin-spin
inverse correlation length at point $h=0$:
\begin{equation}
   \label{eq:kappa11}
   \kappa^{''}_{hh}=2\left\{
   \frac{1}{\lambda_1}\left[\psi_1^{+}V_2^{(1)}\psi_1
   +\sum_{k=1}^{N_2}\frac{(\psi_1^{+}V_1^{(12)}\varphi_k)^2}{\lambda_1
   -\lambda_k^{(2)}}\right]
   -\frac{1}{\lambda_1^{(2)}}\left[\varphi_1^{+}V_2^{(2)}\varphi_1
   +\sum_{k=1}^{N_1}\frac{(\psi_k^{+}V_1^{(12)}\varphi_1)^2}
   {\lambda_1^{(2)}-\lambda_k^{(1)}}\right]
   \right\}\;.
\end{equation}

Calculating the largest eigenvalue $\Lambda_1$ up to the terms
of fourth order in $h$, we find the following result (suited for
programming) for the initial nonlinear susceptibility:
\begin{mathletters}
\label{eq:chi4}
\begin{equation}
   \chi_n^{(4)}=\frac{12}{n^2\lambda_1}\left[\frac{1}{\lambda_1}Q^2
   -2(Q_1+Q_2+Q_3+Q_4+Q_5-Q_6+Q_7-Q_8)\right]
\end{equation}
with
\begin{eqnarray}
   &&Q=\psi_1^{+}V_2^{(1)}\psi_1
   +\sum_{k=1}^{N_2}\frac{(\psi_1^{+}V_1^{(12)}\varphi_k)^2}{\lambda_1
   -\lambda_k^{(2)}}\;,
   \nonumber\\ \nonumber\\
   &&Q_1=\psi_1^{+}V_4^{(1)}\psi_1\;,\qquad
   Q_2=\sum_{k=2}^{N_2}\frac{(\psi_1^{+}V_2^{(1)}\psi_k)^2}{\lambda_1
   -\lambda_k^{(1)}}\;,\qquad
   Q_3=2\sum_{k=1}^{N_2}\frac{
   (\psi_1^{+}V_1^{(12)}\varphi_k)(\psi_1^{+}V_3^{(12)}\varphi_k)}
   {\lambda_1-\lambda_k^{(2)}}\;,
   \nonumber\\ \nonumber\\
   &&Q_4=2\sum_{k=1}^{N_2}\sum_{l=2}^{N_1}\frac{
   (\psi_1^{+}V_1^{(12)}\varphi_k)(\varphi_k^{+}V_1^{(21)}\psi_l)
   (\psi_l^{+}V_2^{(1)}\psi_1)}
   {(\lambda_1-\lambda_k^{(2)})(\lambda_1-\lambda_l^{(1)})}\;,
   \nonumber\\ \nonumber\\
   &&Q_5=\sum_{k=1}^{N_2}\sum_{l=1}^{N_2}\frac{
   (\psi_1^{+}V_1^{(12)}\varphi_k)(\varphi_k^{+}V_2^{(2)}\varphi_l)
   (\varphi_l^{+}V_1^{(21)}\psi_1)}
   {(\lambda_1-\lambda_k^{(2)})(\lambda_1-\lambda_l^{(2)})}\;,
   \\
   &&Q_6=\psi_1^{+}V_2^{(1)}\psi_1
   \sum_{k=1}^{N_2}\frac{(\psi_1^{+}V_1^{(12)}\varphi_k)^2}
   {(\lambda_1-\lambda_k^{(2)})^2}\;,
   \nonumber\\ \nonumber\\
   &&Q_7=\sum_{k=1}^{N_2}\sum_{l=2}^{N_1}\sum_{m=1}^{N_2}\frac{
   (\psi_1^{+}V_1^{(12)}\varphi_k)(\varphi_k^{+}V_1^{(21)}\psi_l)
   (\psi_l^{+}V_1^{(12)}\varphi_m)(\varphi_m^{+}V_1^{(21)}\psi_1)}
   {(\lambda_1-\lambda_k^{(2)})(\lambda_1-\lambda_l^{(1)})
   (\lambda_1-\lambda_m^{(2)})}\;,
   \nonumber\\ \nonumber\\
   &&Q_8=
   \sum_{k=1}^{N_2}\frac{(\psi_1^{+}V_1^{(12)}\varphi_k)^2}
   {(\lambda_1-\lambda_k^{(2)})^2}
   \sum_{l=1}^{N_2}\frac{(\psi_1^{+}V_1^{(12)}\varphi_l)^2}
   {\lambda_1-\lambda_l^{(2)}}\;.
   \nonumber
\end{eqnarray}
\end{mathletters}

Explicit expressions for the matrix elements of subblocks
$V_1^{(12)}$, $V_2^{(1)}$, $V_2^{(2)}$, $V_3^{(12)}$, and
$V_4^{(1)}$ are given in the Appendix.
Notice that the full eigenproblems for the matrices $V^{(1)}$ and
$V^{(2)}$ were solved by means of the $C$ library function pair
{\sf tred2} and {\sf tqli} \cite{PTVF}.
All calculations were carried out on a personal computer IBM PC-486
with the operating system FreeBSD.


\section{Behavior of the Critical Amplitude Ratios}
\label{sec:BCAR}
As already noted in Sec.\ \ref{sec:SAPFE}, the FSS amplitude
combinations normally should depend on the interaction ratios
which change the shape of a subsystem.
This is in agreement with basic ideas of the renormalization-group
theory.

We will consider how the amplitude combinations behave versus the
interaction anisotropy parameters in the three-dimensional Ising
model.
To eliminate as effectively as possible nonuniversal quantities from
the critical FSS amplitude combinations, it is naturally to consider
the behavior of such combinations which do not include in its
composition the nonuniversal metric and geometry factors.
In other words, it is reasonable to base again on the
Privman-Fisher equations in choosing the combinations.

For a cluster $n\times n\times\infty$, the values of linear
and nonlinear susceptibilities and derivative $\kappa_{hh}^{''}$
are taken at critical temperatures which were determined by the
$(n-1,n)$ cluster pair, i.e., through the equation
\begin{equation}
   \label{eq:Tc}
   (n-1)\kappa_{n-1}(T_c^{(n-1,n)})=n\kappa_n(T_c^{(n-1,n)})\;.
\end{equation}
Since we are able to perform the calculations for subsystems with
$n\leq4$, this allows us to have two independent iterations,
with $(2,3)$ and $(3,4)$ pairs (a degenerate pair $(1,2)$ was
eliminated from the consideration due to its anomalies \cite{YuS}).
Two steps of converging iterations give already a possibility
to extract valuable information about tendencies in the change of
quantities with increasing $n$.

The results for the critical temperatures and critical
FSS amplitude ratios $A_{\kappa_{hh}^{''}}/A_\chi$ and
$A_{\chi^(4)}A_s/A_\chi^2$ depending on $J_x/J_z$ and $J_y/J_x$
in the cases of both cluster pairs $(2,3)$ and $(3,4)$ are collected
in Table I.

Let us first discuss the behavior of amplitude ratios against
$J_x/J_z$.
Fig.\,1\ \cite{Figs} shows the dependencies of
$A_{\kappa_{hh}^{''}}/A_\chi$ on $J_x/J_z$ at different fixed
values of $J_y/J_x$.
If $J_y/J_x=0$, the bar $n\times n\times\infty$ is reduced to
statistically independent strips.
In two-dimensional space, the dependence on an anisotropy parameter
must be, as it follows from Eqs.\,(\ref{eq:kappa_n})
and (\ref{eq:f_n}), absent.
The $J_y/J_x=0$ plot given in Fig.\,1 confirms this: in the case of
$(3,4)$ approximation, the ratio $A_{\kappa_{hh}^{''}}/A_\chi$ is
unchanged within the relative root-mean-square error $0.08\%$.
We expect the constancy of amplitude ratios also at $J_y/J_x=1$.
The corresponding line shown in Fig.\,1 can be considered as a
straight one with an accuracy $0.31\%$.
Importantly, the analogous error for the $(2,3)$ pair is $0.45\%$.
Thus, the deviations versus $J_x/J_z$ fall as $n$ increases.

For intermediate fixed values of $J_y/J_x$, the amplitude ratio has
weak but still quite detectible dependence against $J_x/J_z$.
As it is seen in Fig.\,1, variances are most essential in the range
$10^{-1}\alt J_y/J_x\leq1$.
Now the amplitude ratio dependence does not tend to disappear with
increasing $n$.
For example, when $J_y/J_x=0.1$, the mean values and the
root-mean-square uncertainties (the latter are shown in parentheses)
for the quantity $A_{\kappa_{hh}^{''}}/A_\chi$ in the range
$10^{-3}\le J_x/J_z\le1$ equal $1.37(2)$ and $1.35(2)$ for the
pairs $(2,3)$ and $(3,4)$, respectively.

Similar picture is observed for the
$A_{\chi^(4)}A_s/A_\chi^2$ (see Table I) and
for the $A_s/A_f$ (Table I of Ref.\ I).
As a whole one can conclude that the critical FSS amplitude
combinations which do not contain the nonuniversal factors $C_1$,
$C_2$, and $G$ display a tendency to the universality under the
anisotropy parameter $J_x/J_z$, when this is small (i.e., by a
quasi-one-dimensional nature of interactions in the system).

We discuss now the behavior of the amplitude ratios as a function of
the second anisotropy parameter, namely $J_y/J_x$.
In I the amplitudes of the inverse correlation lengths and the free
energy have been calculated for $J_y/J_x\in[0,1]$.
The amplitude $A_f$ was found from a set of two equations
\begin{equation}
   \label{eq:Af}
   f_n=f_\infty+n^{-d}A_f
\end{equation}
with $n=3$ and 4 [``background'' $f_\infty$ is a second unknown
variable in Eq.\,(\ref{eq:Af})].
In this paper we prolong such calculations to $J_y/J_x=4$ and
use not only the $(3,4)$ pair but also the $(2,3)$ one, to observe
the evolution with increasing $n$.
The results are shown in Fig.\,2.
It can be seen from the figure that when the parameter $J_y/J_x$
increases from zero, the ratio $A_s/A_f$ first increases
monotonically from some finite value which
decreases with growth of cluster size.
Then the $A_s/A_f$ attains a smooth maximum near $J_y/J_x=1$.
Lastly, the ratio $A_s/A_f$ falls monotonically as $J_y/J_x$
becomes large.

It follows from the obtained data that the values of $A_s/A_f$ are
equal between themselves for $J_y/J_x$ and $(J_y/J_x)^{-1}$ and
this is better the smaller $J_x/J_z$.
Clearly, the source of the $J_y/J_x\leftrightarrow (J_y/J_x)^{-1}$
symmetry is caused by the (approximate) $J_x/J_z$ independency of
amplitude ratios.
The inversion symmetry leads in turn to an existence of extremum
at $J_y/J_x=1$.
Inasmuch as the extremum (maximum) is smooth, there is a local
universality under the second anisotropy parameter $J_y/J_x$.
We conclude that in the domain
$0<J_x/J_z\ll1$ and $|J_y/J_x-1|\ll1$ there exist a complete
universality of the amplitude ratio $A_s/A_f$.
Note also that in the maximum region both curves go quite near
one to the other; such neighborhood characterizes the arrived
convergence of $(2,3)$ and $(3,4)$ approximations.

We make the quantitative comparison of our data in the two-dimensional
case.
Space dimensionality $d$ changes discontinuously from 3 to 2 in the
limit $J_y/J_x\to0$.
This leads in turn to a finite jump in values of $A_s/A_f$ at
$J_y/J_x=0^{+}$ and $J_y/J_x=0$.
Using Eq.\,(\ref{eq:Af}) one finds for the $(n-1,n)$ pair that
\begin{equation}
   \label{eq:AsAf_0}
   (A_s/A_f)_0=\frac{n(n-1)(2n-1)}{3n(n-1)+1}(A_s/A_f)_{0^{+}}\;,
\end{equation}
where the subscripts $0$ and $0^{+}$ denote the values at
$J_y/J_x=0$ and $0^{+}$, respectively.
Hence, recalculating the values at $J_y/J_x=0^{+}$ to
the values at $J_y/J_x=0$, we obtain $(A_s/A_f)_0=3.0143$ for the
$(2,3)$ pair in the case $J_x/J_z=10^{-3}$.
For the $(3,4)$ pair at the same value of $J_x/J_z$,
$(A_s/A_f)_0=3.0087$.
In the two-dimensional Ising lattice, the exact value $A_s/A_f=3$
(in Fig.\,2, it is shown by the open circle on the ordinate axis).
Consequently, the error is reduced from $0.48\%$ to $0.29\%$ by
going from $(2,3)$ to $(3,4)$ pair.

Qualitatively similar picture takes place for the ratios
$A_{\kappa_{hh}^{''}}/A_\chi$ and $A_{\chi^{(4)}}A_s/A_{\chi}^2$
(Figs.\,3 and 4).
Here again the amplitude ratios have finite values at
$J_y/J_x=0^{+}$ which vanish with increasing $n$ according to the
$1/n$ law.
There is again a monotonic increase for small values of $J_y/J_x$, a
broad maximum near $J_y/J_x=1$ and then a monotonic decrease
by large $J_y/J_x$.
The inversion symmetry
$J_y/J_x\leftrightarrow (J_y/J_x)^{-1}$ applies again.

Let us compare the results for the discussed amplitude ratios
in the two-dimensional limit.
Since
\begin{equation}
   \label{eq:A1_0}
   (A_{\kappa_{hh}^{''}}/A_\chi)_0
   =n(A_{\kappa_{hh}^{''}}/A_\chi)_{0^{+}}\;,
\end{equation}
taking from Table I the values at $J_y/J_x=0^{+}$, one finds that
$(A_{\kappa_{hh}^{''}}/A_\chi)_0$ equals 1.959 and 1.962 for the
$(2,3)$ and $(3,4)$ approximations, respectively; (the point
$J_x/J_z=10^{-3}$ has been used).
These values are in good agreement with available estimates of
the amplitude ratio for the two-dimensional isotropic Ising models,
$A_{\kappa_{hh}^{''}}/A_\chi=1.95-1.96$ (Ref.\cite{BN}).

One can argue analogously for the combination
$A_{\chi^{(4)}}A_s/A_{\chi}^2$,
calculation being carried out by the formula
\begin{equation}
   \label{eq:A2_0}
   (A_{\chi^{(4)}}A_s/A_{\chi}^2)_0
   =n(A_{\chi^{(4)}}A_s/A_{\chi}^2)_{0^{+}}\;.
\end{equation}
Taking from Table I the necessary data, we find that at
$J_x/J_z=10^{-3}$ the ratio $(A_{\chi^{(4)}}A_s/A_{\chi}^2)_0=5.7892$
and 5.7976 for the pairs $(2,3)$ and $(3,4)$, respectively.
These results are in excellent agreement with the value
$A_{\chi^{(4)}}A_s/A_{\chi}^2=5.797\,28(5)$ following from the
calculations \cite{BD}.

We discuss now a reason which could lead to a disappearance of the
parameter $J_x/J_z$ from scaling functions for the system with
the dominant intrachain interaction ($J_z\gg J_x,J_y$).
The bulk correlation-length amplitude is a function of the
interaction constants.
Let $\xi_0^{(x)}=\phi(J_x/k_BT_c,J_y/k_BT_c,J_z/k_BT_c)$.
In a lattice which is infinite in all three directions, the
function $\phi(x,y,z)$ is symmetrical under the replacement of its 2nd
and 3rd arguments: $\phi(x,y,z)=\phi(x,z,y)$.
Moreover, $\phi(0,y,z)=0$ which means the absence of correlations
in the direction along which there are no interactions.
Let the expansion on the first argument begin from a term $x$ in
some power $q$.
In accordance with physical reasoning, it is apparent that in the
infinite lattice the amplitude $\xi_0^{(y)}$ has the same functional
form as $\xi_0^{(x)}$ but, of course, with replaced arguments:
$\xi_0^{(y)}=\phi(J_y/k_BT_c,J_x/k_BT_c,J_z/k_BT_c)$.
Further, the critical temperature of a quasi-one-dimensional Ising
model is given by \cite{WGFF}
\begin{equation}
   \label{eq:TcQ1D}
   k_BT_c/J_z=2\left[\ln\left(\frac{J_z}{J_x+J_y}\right)
   -\ln\ln\left(\frac{J_z}{J_x+J_y}\right)+O(1)\right]^{-1}\;.
\end{equation}
Due to the fact that a decrease of $k_BT_c/J_z$ is logarithmically
slow, the arguments of function
$\phi$ $J_x/k_BT_c=J_x/J_z\cdot J_z/k_BT_c$ and
$J_y/k_BT_c=J_y/J_x\cdot J_x/J_z\cdot J_z/k_BT_c$
are small for small $J_x/J_z$.
Expanding the bulk correlation-length amplitudes in Taylor series
and keeping only the leading asymptotical term, we obtain that as
$J_x/J_z\to0$ the aspect ratio $a_y/a_x\simeq(J_y/J_x)^{-q}$.
The anisotropy parameter $J_x/J_z$ did drop out.

In the light of the above statements the Privman-Fisher equations for
the quasi-one-dimensional system ($J_x,J_y\ll J_z$) can be written as
\begin{equation}
   \label{eq:kappa_Q1D}
   \kappa_n(t, h)
   = n^{-1}G(J_x/J_z,J_y/J_x)
   X(C_1tn^{y_T},C_2hn^{y_h};J_y/J_x)
\end{equation}
and
\begin{equation}
   \label{eq:f_Q1D}
   f_n^{(s)}(t, h)
   = n^{-d}G(J_x/J_z,J_y/J_x)
   Y(C_1tn^{y_T},C_2hn^{y_h};J_y/J_x)\;,
\end{equation}
where $d=|{\rm sign}(J_x)|+|{\rm sign}(J_y)|+|{\rm sign}(J_z)|$.
The geometry prefactor is normalized so that
$G(1,1)=1$ and the dependence $G(J_x/J_z,0)$ is given by
Eqs.\,(\ref{eq:gamma_2D}) and (\ref{eq:Tc2D}).

We know that the $J_y/J_x\leftrightarrow(J_y/J_x)^{-1}$
invariance and the analyticity of scaling functions leads to the
existence of a smooth extremum for the critical amplitude ratios
of the $n\times n\times\infty$ parallelepipeds at $J_y/J_x=1$.
In this context note that in the case of parallelepipeds with a
{\it rectangular\/} cross section, $n_x\times n_y\times\infty$,
one should expect the extrema for the certain ratios of critical
amplitudes at $J_y/J_x=(n_y/n_x)^{1/q}$.


\section{Completely Isotropic Lattice}
\label{sec:CIL}
At present, the critical point of the fully isotropic
simple-cubic Ising lattice is located to a high degree of accuracy:
$K_c=0.221\,655(1)$ (Ref.\cite{GT} and references therein).
It is known also with large accuracy the free energy at criticality:
$f_\infty=0.777\,90(2)$, Ref\cite{M}.
On the other hand, we can carry out the calculations for subsystems
$n\times n\times\infty$
with three numbers of lattice layers $n=2$, 3, and 4.
This allows, generally speaking, to perform the three-point
extrapolations for accelerating the convergence and to improve the
estimates of the universal critical amplitudes $A_s$ and $A_f$ and
also the universal critical amplitude ratios
$A_{\kappa_{hh}^{''}}/A_\chi$ and $A_{\chi^{(4)}}/A_\chi^2$.
The results are summarized in Table II.

One can compare our results with the available estimates.
According to a Monte Carlo simulation on periodic cylinders
$n\times n\times128$ with $n=4$, 6, 8, and 10 (Ref.\cite{W}), the
critical FSS amplitude of the spin-spin inverse correlation length
equals $A_s=1.30(3)$.
Our estimate $A_s=1.26(5)$ is consistent with the one above.

Information concerning the absolute amplitude of the free energy can
be extracted from the published data by the indirect route.
Indeed, in accordance with the calculations \cite{H} carried out in the
quantum limit of a three-dimensional Ising model,
$A_s/A_f=3.671(6)$.
Using the above estimate $A_s=1.30(3)$ one finds $A_f=0.354(9)$.
The value $A_f=0.37(3)$ given in Table II agrees with this estimate.
Note that the critical FSS free-energy amplitude of the
three-dimensional Ising lattice with the shape of periodic cubes is
equal to $A_f^{cube}=0.625(5)$, Ref.\cite{M}.
For a comparison note also that in the two-dimensional space the
free-energy amplitudes of periodic Ising strips and squares are equal
correspondingly to $\pi/12=0.261\,799$ and
$\ln(2^{1/4}+2^{-1/2})=0.639\,911$ (see, e.g., Ref.\cite{PPHA}).

We do not know from the literature any estimates for the universal
ratio $A_{\kappa_{hh}^{''}}/A_\chi$ in three dimensions.
One can note that out of all the four quantities discussed
in this section, the combination $A_{\kappa_{hh}^{''}}/A_\chi$ has the
best convergence in $n$.

Finally, for the universal combination
$A_{\chi^{(4)}}/A_\chi^2$ which is the finite-size cumulant ratio
${\bar g}_\infty$, there exists an estimate only in the order of
magnitude: ${\bar g}_\infty\sim3$ (Ref.\cite{SD}, see also the reviews
\cite{PPHA}).
We have succeeded in obtaining this quantity to an accuracy of $5\%$.


\section{Conclusions}
\label{sec:Con}
In this article, the author has presented large-scale transfer-matrix
calculations for different finite-size amplitudes of a fully
anisotropic simple-cubic Ising lattice in the shape of
$n\times n\times\infty$ bars with periodic boundaries in both
transverse directions.
The behavior of amplitude combinations (ratios) which do not contain
the nonuniversal metric factors and geometry prefactor was studied
depending on the interaction anisotropy parameters
$J_x/J_z$ and $J_y/J_x$.
It has been established that these amplitude ratios practically cease
to depend on the anisotropy parameter $J_x/J_z\to0$ and, what
more, this is true for wide interval, $J_x/J_z\alt10^{-1}$.

It was shown that as a function of $J_y/J_x$ the amplitude ratios
have a smooth extremum (maximum) near $J_y/J_x=1$.
As a result, the critical finite-size amplitude combinations are
universal with respect to both anisotropy parameters in the domain
$0<J_x/J_z\ll1$ and $|J_y/J_x-1|\ll1$.

A mechanism leading to the $J_x/J_z$ independence of certain
amplitude ratios was proposed.
By this the $J_x\leftrightarrow J_y$ invariance together with the
analyticity of the scaling functions explains the existence of a
smooth extremum in the amplitude combinations at $J_y/J_x=1$.

In the case of fully isotropic interactions ($J_x=J_y=J_z$) for
which the high accuracy values of critical coupling and critical
free energy are available, the better estimates have been found for
the universal critical finite-size amplitudes of the spin-spin
inverse correlation length and singular part of the free energy per
site, as well as for the universal amplitude ratios
$A_{\kappa_{hh}^{''}}/A_\chi$ and $A_{\chi^{(4)}}/A_\chi^2$.


\section*{acknowledgment}
It is a pleasure to thank the referee for instructive remarks.


\appendix
\section*{Matrix Elements of $V_1$, $V_2$, $V_3$, and $V_4$ in the
quasidiagonal representation of $V$}
\label{sec:App}
The sizes of upper and lower subblocks in the expansion
(\ref{eq:tilde_V}) of TM for a cylinder $n\times n\times\infty$ are
equal between themselves ($N_1=N_2$) by odd $n$ and unequal
($N_1\ne N_2$) by even $n$.
Consider the cases of even and odd $n$ separately.

Matrix elements of subblocks $V^{(1)}$ and $V^{(2)}$ for the bars
$3\times 3\times\infty$ and $4\times 4\times\infty$ have been done
in \cite{Yu93} and I.
For the $2\times 2\times\infty$ cluster, subblocks $V^{(1)}$ and
$V^{(2)}$ have, respectively, the sizes $5\times5$ and $2\times2$
\cite{Yu93}.
As for the $4\times4\times\infty$ cluster, their matrix elements
are given by
\begin{mathletters}
\label{eq:A_V1V2}
\begin{equation}
   \label{eq:A_V1}
   V_{ij}^{(1)}=A_{ij}G_{ij}
\end{equation}
and
\begin{equation}
   \label{eq:A_V2}
   V_{ij}^{(2)}=A_{ij}{\tilde G}_{ij}\;,
\end{equation}
\end{mathletters}
where
\begin{equation}
   \label{eq:A_Aij}
   A_{ij}=\frac{\max(n_i,n_j)}{\sqrt{n_in_j}}\exp\left[
   \case{1}{2}(m_i^a+m_j^a)K_x+\case{1}{2}(m_i^b+m_j^b)K_y\right]\;,
\end{equation}
\begin{mathletters}
\label{eq:A_Gij}
\begin{equation}
   G_{ij}=g_0^{(ij)}+2\sum_{s=1}^{\case{1}{2}n^2}g_s^{(ij)}
   \cosh(2sK_z)\;,
\end{equation}
and
\begin{equation}
   {\tilde G}_{ij}=2\sum_{s=1}^{\case{1}{2}n^2}{\tilde g}_s^{(ij)}
   \sinh(2sK_z)\;.
\end{equation}
\end{mathletters}
The coefficients $n_i$, $m_i^a$, $m_i^b$, and $m_i$ for the periodic
cylinder $2\times2\times\infty$ are
\begin{equation}
   \label{eq:A_nm}
   n_i=\{2,8,2,2,2\},\quad
   m_i^a=\{4,0,-4,-4,4\},\quad
   m_i^b=\{4,0,4,-4,-4\},\quad
   m_i=\{4,2,0,0,0\}.
\end{equation}
As in the case of $4\times4\times\infty$ cluster, the coefficients
$g_s^{(ij)}$ satisfy again the condition
\begin{equation}
   \label{eq:A_gij}
   g_0^{(ij)}+2\sum_{s=1}^{\case{1}{2}n^2}g_s^{(ij)}
   =\min(n_i,n_j)\;.
\end{equation}
For $g_s^{(ij)}$ with $s\ne0$ one has
\begin{eqnarray}
   \label{eq:A_gij_s}
   &&11)\;0\ 1\quad 21)\;1\ 0\quad 22)\;0\ 1\quad 31)\;0\ 0\quad
   32)\;1\ 0\quad 33)\;0\ 1\quad 41)\;0\ 0\quad 42)\;1\ 0\quad
   \nonumber\\
   &&43)\;0\ 0\quad 44)\;0\ 1\quad 51)\;0\ 0\quad 52)\;1\ 0\quad
   53)\;0\ 0\quad 54)\;0\ 0\quad 55)\;0\ 1\;.
\end{eqnarray}
Finally, the ${\tilde g}_s^{(ij)}$ coefficients of a cluster
$2\times2\times\infty$ equal
\begin{equation}
   \label{eq:A_tilde_gij_s}
   11)\;0\ 1\quad 21)\;1\ 0\quad 22)\;0\ 1\;.
\end{equation}

Taking (from \cite{Yu93} and I) the basis functions of irreducible
representations we find
the matrix elements of subblocks $V_1^{(12)}$ for bars
$n\times n\times\infty$ with even $n$ (i.e., for cylinders
$2\times2\times\infty$ and $4\times4\times\infty$):
\begin{equation}
   \label{eq:A_V1ij}
   (V_1^{(12)})_{ij}=\case{1}{2}A_{ij}(m_jG_{ij}
   +m_i{\tilde G}_{ij})\;,
\end{equation}
where $m_i=0$ by $i>N_2$; here and below we regard, for
definiteness sake, ${\tilde G}_{ij}=0$ when $i$ or $j>N_2$.
Analogously one evaluates for subblocks $V_2^{(1)}$ and $V_2^{(2)}$:
\begin{mathletters}
\label{eq:A_V2ij}
\begin{equation}
   (V_2^{(1)})_{ij}=\frac{1}{2^22!}A_{ij}[(m_i^2+m_j^2)G_{ij}
   +2m_im_j{\tilde G}_{ij}]
\end{equation}
and
\begin{equation}
   (V_2^{(2)})_{ij}=\frac{1}{2^22!}
   A_{ij}[(m_i^2+m_j^2){\tilde G}_{ij}+2m_im_jG_{ij}]\;.
\end{equation}
\end{mathletters}
The matrix elements of subblock $V_3^{(12)}$ are
\begin{equation}
   \label{eq:A_V3ij}
   (V_3^{(12)})_{ij}=\frac{1}{2^33!}
   A_{ij}[m_j(3m_i^2+m_j^2)G_{ij}
   +m_i(m_i^2+3m_j^2){\tilde G}_{ij}]\;.
\end{equation}
Lastly, for subblock $V_4^{(1)}$ we find
\begin{equation}
   \label{eq:A_V4ij}
   (V_4^{(1)})_{ij}=\frac{1}{2^44!}
   A_{ij}[(m_i^4+6m_i^2m_j^2+m_j^4)G_{ij}
   +4m_im_j(m_i^2+m_j^2){\tilde G}_{ij}]\;.
\end{equation}

In the case of a cluster $n\times n\times\infty$ with odd number
of layers $n$, the formulas are more uniform because the same set of
coefficients $g_s^{(ij)}$ enters into matrix elements for subblocks
of both representations.
Introducing the auxiliary quantities
\begin{mathletters}
\label{eq:A_Hij}
\begin{equation}
   H^{(+)}_{ij}=\sum_{s=1}^{\case{1}{2}(n^2+1)}
   |g_s^{(ij)}|\exp[(2s-1)K_z{\rm sign}(g_s^{(ij)})]
\end{equation}
and
\begin{equation}
   H^{(-)}_{ij}=\sum_{s=1}^{\case{1}{2}(n^2+1)}
   |g_s^{(ij)}|\exp[-(2s-1)K_z{\rm sign}(g_s^{(ij)})]\;,
\end{equation}
\end{mathletters}
we obtain for even ($k$) terms of transfer-matrix expansion:
\begin{mathletters}
\label{eq:A_Vkij}
\begin{equation}
   (V_k^{(1)})_{ij}=\frac{1}{2^kk!}A_{ij}
   \left[(m_i+m_j)^{k}H_{ij}^{(+)}+(m_i-m_j)^kH_{ij}^{(-)}\right]
\end{equation}
and
\begin{equation}
   \label{eq:A_Vkij_b}
   (V_k^{(2)})_{ij}=\frac{1}{2^kk!}A_{ij}
   \left[(m_i+m_j)^{k}H_{ij}^{(+)}-(m_i-m_j)^kH_{ij}^{(-)}\right]\;,
\end{equation}
\end{mathletters}
where $A_{ij}$ equal (\ref{eq:A_Aij}) but, of course, with their sets
of coefficients.
By odd $k$, i.e.\ for subblocks $V_k^{(12)}$, the expressions for
the matrix elements have the form
\begin{equation}
   \label{eq:A_Vkij_odd}
   (V_k^{(12)})_{ij}=\frac{1}{2^kk!}A_{ij}
   \left[(m_i+m_j)^{k}H_{ij}^{(+)}-(m_i-m_j)^kH_{ij}^{(-)}\right]\;.
\end{equation}



\begin{table}
\label{tab:A1A2}
\caption{
Critical-point amplitude ratios
$A_{\kappa_{hh}^{''}}/A_\chi$ and $A_{\chi^{(4)}}A_s/A_\chi^2$ as a
function of anisotropy parameters $J_x/J_z$ and $J_y/J_x$.
Approximations by cluster pairs $(2,3)$ and $(3,4)$.}
\begin{tabular}{llcccccc}
&&&$(2,3)$ pair&&&$(3,4)$ pair&\\[-4mm]
$J_y/J_x$&$J_x/J_z$&
\\[-3.5mm] \cline{3-5} \cline{6-8}
&&$k_BT_c/J_z$&$A_{\kappa_{hh}^{''}}/A_\chi$&
$A_{\chi^{(4)}}A_s/A_\chi^2$&
$k_BT_c/J_z$&$A_{\kappa_{hh}^{''}}/A_\chi$&
$A_{\chi^{(4)}}A_s/A_\chi^2$\\[1mm]
\tableline
$0^+$&1.0  &2.367640&0.65214&1.92695&2.320811&0.48946&1.44560\\
     &0.5  &1.699790&0.65173&1.92388&1.662411&0.48991&1.44730\\
     &0.1  &0.921068&0.65279&1.92832&0.910794&0.49034&1.44919\\
     &0.01 &0.513815&0.65309&1.92968&0.510586&0.49039&1.44940\\
     &0.001&0.345015&0.65310&1.92972&0.343461&0.49039&1.44940\\ \\

 0.1 &1.0  &2.845802&1.34451&3.60933&2.817633&1.32110&3.48345\\
     &0.5  &1.979483&1.35368&3.62760&1.959011&1.34212&3.54079\\
     &0.1  &1.024601&1.38006&3.70054&1.019932&1.35995&3.59108\\
     &0.01 &0.550499&1.38661&3.71950&0.549216&1.36225&3.59760\\
     &0.001&0.362466&1.38685&3.72021&0.361871&1.36233&3.59781\\ \\

 0.5 &1.0  &3.819394&1.65412&4.55222&3.739735&1.67447&4.59324\\
     &0.5  &2.533705&1.65621&4.54634&2.487136&1.68196&4.61381\\
     &0.1  &1.211125&1.67095&4.59073&1.199032&1.68882&4.63492\\
     &0.01 &0.611620&1.67545&4.60591&0.608159&1.68976&4.63782\\
     &0.001&0.390357&1.67564&4.60656&0.388821&1.68979&4.63793\\ \\

 1.0 &1.0  &4.685960&1.68710&4.67454&4.581044&1.70507&4.70713\\
     &0.5  &3.024529&1.68407&4.64512&2.960469&1.71114&4.72047\\
     &0.1  &1.366300&1.69711&4.68102&1.350375&1.71743&4.73841\\
     &0.01 &0.658794&1.70231&4.69869&0.654587&1.71832&4.74101\\
     &0.001&0.410971&1.70256&4.69953&0.409173&1.71835&4.74110\\ \\

 1.5 &1.0  &5.418190&1.68220&4.66797&5.310630&1.69188&4.66481\\
     &0.5  &3.442977&1.67334&4.61261&3.368233&1.69831&4.67598\\
     &0.1  &1.494583&1.68607&4.64298&1.475982&1.70726&4.70162\\
     &0.01 &0.695761&1.69307&4.66651&0.691216&1.70870&4.70596\\
     &0.001&0.426622&1.69344&4.66777&0.424737&1.70875&4.70612\\ \\

 2.0 &1.0  &6.068336&1.66945&4.63471&5.975643&1.66826&4.58676\\
     &0.5  &3.819394&1.65412&4.55222&3.739735&1.67447&4.59324\\
     &0.1  &1.608392&1.66577&4.57383&1.587593&1.68729&4.63016\\
     &0.01 &0.727256&1.67510&4.60471&0.722535&1.68970&4.63763\\
     &0.001&0.439633&1.67563&4.60652&0.437724&1.68979&4.63792\\ \\
\end{tabular}
\end{table}


\newpage
\begin{table}
\label{tab:Iso}
\caption{
Estimates of the universal critical amplitude combinations
for the fully isotropic three-dimensional simple-cubic Ising lattice
in the geometry of infinitely long cyclic parallelepipeds with the
square cross section.
For all quantities, with the exception of $A_f$, the three-point
extrapolations were performed by the Shanks' transform.$^{24}$
In the case of $A_f$ as a realistic estimate, the last term of finite
sequence has been taken.}
\begin{tabular}{ccccc}
$n$&$A_s$&$A_f$&
$A_{\kappa_{hh}^{\prime\prime}}/A_\chi$&
$A_{\chi^{(4)}}/A^2_\chi$\\[2mm]
\tableline
 2       &1.458347 &0.451886 &1.784594 &3.497834\\
 3       &1.353169 &0.409959 &1.763848 &3.661609\\
 4       &1.302661 &0.378507 &1.755304 &3.759097\\[2mm]
$\infty$ &1.26(5)\ \ \ &0.37(3)\ \ \ &1.749(6) &3.9(2)\ \ \ \ \\
\end{tabular}
\end{table}


\begin{figure}
\label{fig:A1_JxJz}
\caption{Amplitude ratio $A_{\kappa_{hh}^{''}}/A_\chi$ vs the
anisotropy parameter $J_x/J_z$ by different fixed values of
$J_y/J_x$. Periodic cylinder $4\times4\times\infty$ at critical
temperatures $T_c^{(3,4)}$.}
\end{figure}

\begin{figure}
\label{fig:AsAf_JyJx}
\caption{Amplitude ratios $A_s/A_f$ (averaged over
$J_x/J_z=\{1,10^{-1},10^{-2},10^{-3}\}$) as a function of $J_y/J_x$.
Shown is are the approximations by $(2,3)$ and $(3,4)$ cluster
pairs --- dashed and solid lines, respectively.
Open circle on the ordinate axis is a value of $A_s/A_f (=3)$ for
the two-dimensional Ising model.}
\end{figure}

\begin{figure}
\label{fig:A1_JyJx}
\caption{Average amplitude ratio $A_{\kappa_{hh}^{''}}/A_\chi$ vs
$J_y/J_x$; periodic cylinder $4\times4\times\infty$ at critical
temperatures $T_c^{(3,4)}$.
Full circle on the ordinate axis corresponds to the
$A_{\kappa_{hh}^{''}}/A_\chi$ at $J_y/J_x=0^{+}$, while open one is
a value of that ratio in the two-dimensional Ising lattice.}
\end{figure}

\begin{figure}
\label{fig:A2_JyJx}
\caption{The same as in Fig.\,3 but for the amplitude ratio
$A_{\chi^{(4)}}A_s/A^2_\chi$.
\qquad\qquad\qquad\qquad\qquad\qquad}
\end{figure}



\begin{references}
\bibitem{Figs}
All figures were plotted using not only the data presented in
Tables I of this paper and the paper \cite{Yu94}
but there were taken also extra points.
\end{references}
\end{document}